\documentclass[aps,prd,onecolumn,11pt,tightenlines,amsmath,amssymb,superscriptaddress,longbibliography,floatfix]{revtex4-2}
\usepackage{graphicx,bm,booktabs}
\usepackage{xcolor}
\usepackage[colorlinks=true,breaklinks=true,citecolor=black,linkcolor=black,urlcolor=black]{hyperref}
\newcommand{\dd}{\mathrm d}
\newcommand{\ii}{\mathrm i}
\begin{document}
\raggedbottom
\title{Fast finite-width CMB B modes from narrow primordial spectra}
\author{Yuki Inoue}
\thanks{Corresponding author: iyuki@ncu.edu.tw}
\affiliation{Department of Physics, National Central University, Taoyuan, Taiwan}
\affiliation{Center for High Energy and High Field (CHiP), National Central University, Taoyuan, Taiwan}
\affiliation{Institute of Physics, Academia Sinica, Taipei, Taiwan}
\affiliation{Institute of Particle and Nuclear Studies, High Energy Accelerator Research Organization (KEK), Tsukuba, Japan}

\begin{abstract}
{Direct calculation of the cosmic microwave background (CMB) $B$-mode spectrum sourced by high-wavenumber primordial gravitational waves requires resolving rapidly oscillating tensor evolution and projection kernels, making extensions to higher wavenumbers computationally expensive.
We derive analytic expressions for the response to a narrow primordial band using a thin-recombination-layer approximation and the Wentzel--Kramers--Brillouin (WKB) approximation, obtaining the low-multipole scaling $D_\ell\propto\ell(\ell+1)k_c^{-4}$.
We extend this treatment to a semi-analytic calculation that retains the finite width of recombination and the relative phase between the tensor wave and the projection kernel, allowing rapid evaluation at high multipoles.
For five bands centered at $k_c=0.78$--$7.81\,{\rm Mpc}^{-1}$, the spectra agree with refined CLASS calculations to within 3.8\% over $2\leq\ell\leq5000$.
With precomputed backgrounds, single-thread generation takes 2.6--3.2 CPU seconds per band, corresponding to estimated speedups of $1.5\times10^3$--$5.8\times10^4$ relative to CLASS costs inferred from representative subbands.
This reduction in computational cost enables us to extend the spectra to $k_c\simeq140\,{\rm Mpc}^{-1}$, where direct calculations are expensive.
A conditional application to BICEP/Keck data through the 2015 season illustrates how the high-wavenumber extrapolation affects amplitude bounds.}
\end{abstract}
\maketitle

\section{Introduction}
Primordial gravitational waves (GWs) imprint magnetic-type ($B$-mode)
polarization on the cosmic microwave background (CMB)~\cite{Kamionkowski2016}. Localized primordial spectra
require resolving the response to their wavenumber support. At high
comoving wavenumber $k$, direct calculations must handle rapidly
oscillating tensor sources and projection kernels even when the final
angular spectra are smooth. A high-$k$ mode can contribute below its
characteristic angular scale: replacing the projection by a Limber
localization can miss the low-multipole polarization response of a
finite primordial band~\cite{Namikawa2019}.

{Namikawa et al.~\cite{Namikawa2019} restricted their numerical constraints to $k\lesssim10\,{\rm Mpc}^{-1}$ because of the rapid oscillations of the spherical Bessel and source functions.
}

{In a separate study, Ng~\cite{Ng2022} noted that strong cancellations in the projection integral make Boltzmann-code calculations difficult at short wavelengths, and used an analytic treatment of the tensor Sachs--Wolfe line-of-sight integral to extend CMB temperature constraints to higher wavenumbers.
These studies motivate methods that reduce the computational cost of oscillatory integrals while preserving their accuracy.}

Integral reduction has established precedents. The line-of-sight method
separates radial projection from source evolution~\cite{Seljak1996}.
Pritchard and Kamionkowski~\cite{Pritchard2005} develop analytic tensor
polarization, including projection envelopes and recombination-width
damping. Weinberg~\cite{Weinberg2006} replaces a truncated photon hierarchy
by a source integral equation. Flauger and Weinberg~\cite{Flauger2007}
use asymptotic radial kernels and separate rapidly varying sum phases
from difference phases. {These established methods provide the starting point for our calculation.}

Here we specify and test their combination with a local complex photon
response and its first envelope-derivative correction for finite
logarithmic bands. The retained phase measures the mismatch between
tensor evolution and radial projection across the source layer. It
preserves finite-width interference at high multipoles while removing
the need to resolve every wave period. We study the $B$-mode response,
the coefficient and conditions of the simpler closed limit, and serial
cost estimates tied to a stated CLASS reference. A conditional application to BICEP/Keck data through the 2015 season
(BK15)~\cite{BK15,BK15data} tests the propagation to amplitude bounds
and distinguishes an extension to higher $k$ from an improvement in sensitivity.

\section{Finite-band response}
{We assume} linear tensor perturbations in a spatially flat background, equal
primordial power in the two helicities, and Thomson scattering on a fixed
ionization history. {We set} $c=1$ and normalize the scale factor to $a_0=1$.
For conformal time $\eta$, present time $\eta_0$, and multipole $\ell$, {we write}
$\chi=\eta_0-\eta$, $x=k\chi$, and $L=\ell(\ell+1)$. The dimensionless
primordial power $P_t$, summed over helicities, is
{
\begin{equation}
 P_t(k)=\begin{cases}
 \dfrac{A_t}{2\Delta}, & k_-\leq k\leq k_+,\\
 0, & \text{otherwise},
 \end{cases}
 \qquad \Delta=\ln(k_+/k_-).
 \label{eq:band}
\end{equation}
}
{Here $A_t=2\int P_t\dd\ln k$ is the integrated amplitude in our convention, and $k_c=(k_-+k_+)/2$ is the arithmetic band center.}
{We define} the normalized average
$\langle X\rangle_k=\Delta^{-1}\int_{k_-}^{k_+}X\dd\ln k$.

The polarization source $p(k,\eta)$ is per unit primordial tensor mode.
The visibility is $g=q e^{-\tau}$, where $q=a n_e\sigma_T>0$ is the conformal
Thomson rate, $n_e$ is the physical free-electron density, $\sigma_T$ the
Thomson cross section, and $\tau(\eta)=\int_\eta^{\eta_0}q\dd\eta'$.
The exact linear transfer and radial kernel are~\cite{Hu1998,Blas2011}
\begin{align}
 \Delta_\ell^B(k)&=\int_0^{\eta_0}\dd\eta\,g p r_\ell^B(k\chi),\label{eq:los}\\
 r_\ell^B(x)&=\tfrac12[j_\ell'(x)+2j_\ell(x)/x].
 \label{eq:radial}
\end{align}
Primes on the spherical Bessel function $j_\ell$ differentiate $x$.
The source convention is $p=\sqrt6P^{(2)}$, with
$P^{(2)}=(\Theta_2^{(2)}-\sqrt6E_2^{(2)})/10$ the tensor photon quadrupole
combination; $\Theta_2^{(2)}$ and $E_2^{(2)}$ are temperature and
electric-polarization moments~\cite{Hu1998}. An overall source sign cancels
in the $BB$ auto-spectrum. The internal electric-polarization quadrupole
{enters through the scattering source for the $B$-mode response.}

For $C_\ell^{BB}=4\pi\int P_t|\Delta_\ell^B|^2\dd\ln k$,
{we define} $D_\ell^{BB}=LT_{\rm CMB}^2C_\ell^{BB}/(2\pi)$,
using $T_{\rm CMB}=2.7255\times10^6\,\mu{\rm K}$~\cite{Fixsen2009}. Then
\begin{equation}
 D_\ell^{BB}
 =A_tLT_{\rm CMB}^2\langle|\Delta_\ell^B|^2\rangle_k.
 \label{eq:exact}
\end{equation}
No instantaneous-recombination approximation enters this expression.

\section{Local source and finite-width projection}
For modes entering in radiation domination, the short-wavelength metric
transfer is approximated by
\begin{equation}
 T(k,\eta)\simeq\frac{\mathcal B}{k}
    \operatorname{Re}[U(\eta)e^{\ii(k\eta+\varphi)}],
 \label{eq:metric}
\end{equation}
Here $U=a^{-1}$, $\mathcal B=D_\nu H_0\sqrt{\Omega_r}$,
$\varphi(k)$ is a time-independent phase, $H_0$ is the present Hubble
constant {in the $c=1$ convention}, $\Omega_r$ is the radiation density
fraction, and $D_\nu$ is the adopted neutrino amplitude damping~\cite{Weinberg2004}.
{We adopt $H_0=67.36\,\mathrm{km\,s^{-1}\,Mpc^{-1}}$. In the transfer equations, we set $c=1$ and use $H_0/c=2.2469\times10^{-4}\,\mathrm{Mpc}^{-1}$, denoted by $H_0$ in this convention.}
The normalization follows by matching to $T=\sin(k\eta)/(k\eta)$ in
radiation domination, followed by $a^{-1}$ dilution. This assumes
$k\gg k_{\rm eq}$ and a wave period short compared with expansion;
$k_{\rm eq}$ is the horizon wavenumber at matter--radiation equality.
{This metric approximation describes the oscillatory evolution after horizon entry.}

To obtain the photon response, {we use} angular brightness and polarization
variables $F,G$, with tensor azimuthal dependence removed:
\begin{align}
 (\partial_\eta+\ii k\mu+q)F&=T'-qp,\notag\\
 (\partial_\eta+\ii k\mu+q)G&=qp.\label{eq:photons}
\end{align}
{Here $\mu$ is the photon-direction cosine relative to the wavevector,
and time primes denote $\dd/\dd\eta$. The metric derivative drives $F$,
while Thomson scattering couples both variables to the source $p$.
This source is obtained by taking their weighted angular moment,}
\begin{equation*}
 p=\tfrac12\int_{-1}^1[-w_FF+w_GG]\dd\mu.
\end{equation*}
{The two weights specify the angular dependence of the
scattering contribution:}
\begin{align}
 w_F&=\tfrac3{16}(1-\mu^2)^2,\notag\\
 w_G&=\tfrac3{16}(1+6\mu^2+\mu^4).
 \label{eq:weights}
\end{align}
These equations fix the angular
normalization; their relation to the hierarchy and retarded integral
solution is given in the Supplemental Material~\cite{ApproxSupplement}.

We factor out the positive-frequency factor
$(\mathcal B/k)e^{\ii(k\eta+\varphi)}$ and expand the remaining envelopes
in their time derivatives. With
$d=\ii kU+U'$ and $d_\mu=q+\ii k(1+\mu)$, {we define}
\begin{equation}
 J_{F,G}=\tfrac12\int_{-1}^1\frac{w_{F,G}}{d_\mu}\dd\mu,
 \qquad M=1-q(J_F+J_G).
\end{equation}
The leading envelopes are $p_0=-dJ_F/M$,
$F_0=(d-qp_0)/d_\mu$, and $G_0=qp_0/d_\mu$. {The first correction accounts for the time variation of these
leading envelopes. Expanding the inverse operator
$d_\mu+\partial_\eta$ to first order gives}
\begin{equation*}
 p_1=\frac1{2M}\int_{-1}^1
              \frac{w_FF_0'-w_GG_0'}{d_\mu}\dd\mu.
\end{equation*}
{We combine the leading source and this derivative correction
into a complex response normalized by the metric envelope $U$:}
\begin{equation}
 R_{\rm eff}=(p_0+p_1)/U.
 \label{eq:response}
\end{equation}
{Restoring the oscillatory factor then yields the real
polarization source,}
\begin{equation}
 p(k,\eta)\simeq\frac{\mathcal B}{k}
       \operatorname{Re}[UR_{\rm eff}e^{\ii(k\eta+\varphi)}].
 \label{eq:source}
\end{equation}
Thus the time-history integral is replaced by a local periodic response
and its first correction; angular integrals remain numerical. The
complex response retains streaming and scattering phase lag. Its
derivative expansion requires envelopes that vary slowly compared with
$|d_\mu|$ and neglects homogeneous transients. {At the angular endpoint $\mu=-1$, the expansion is controlled by $d_\mu=q$.}
This source approximation is distinct from tight coupling.

In the oscillatory radial domain $x>\sqrt L$, {we use} the
Wentzel--Kramers--Brillouin (WKB) quantities
\begin{align}
 \mu_\ell&=\sqrt{1-L/x^2},\notag\\
 \Phi_\ell&=\sqrt{x^2-L}-\sqrt L\arccos(\sqrt L/x)
                         +\tfrac\pi2(\sqrt L-\ell).
\end{align}
They give $j_\ell\simeq\sin\Phi_\ell/(x\sqrt{\mu_\ell})$. Retaining
phase derivatives in Eq.~\eqref{eq:radial} yields
$r_\ell^B\simeq\mathcal A_B\cos\Phi_\ell$, where
\begin{equation}
 \mathcal A_B=\frac{\sqrt{\mu_\ell}}{2x}.
 \label{eq:amplitudes}
\end{equation}
The WKB derivative diagnostic is $L/(x^3\mu_\ell^3)\ll1$; amplitude
derivatives and extra inverse-$x$ terms are omitted. {The approximation applies away from the turning point $x=\sqrt L$.}

The product of source and projection separates into a sum phase
$\Phi_\ell+x$ and a difference phase $\Psi_\ell=\Phi_\ell-x$.
Their conformal-time rates have magnitudes $k(1+\mu_\ell)$ and
$k(1-\mu_\ell)$, respectively. For $x\gg\sqrt L$,
\begin{equation}
 \Psi_\ell'\simeq\frac{L}{2k\chi^2}.
 \label{eq:mismatch}
\end{equation}
The wave phase and changing propagation distance therefore nearly
cancel in one component. Averaging them independently would discard
this contribution. {After neglecting the rapidly oscillating sum-phase integral,
we express the retained contribution as a coherent integral over the
recombination layer:}
\begin{equation}
 \mathcal J_B(k,\ell)=\int_{\rm rec}\dd\eta\,
         gUR_{\rm eff}\mathcal A_B e^{\ii\Psi_\ell}.
 \label{eq:J}
\end{equation}
{The modulus squared of this integral retains interference
between scattering times. Averaging the overall wave phase over a
sufficiently broad $k$ band and substituting into
Eq.~\eqref{eq:exact} gives}
\begin{equation}
 D_\ell^{BB}\simeq
 \frac{A_tLT_{\rm CMB}^2\mathcal B^2}{8}
     \langle k^{-2}|\mathcal J_B|^2\rangle_k.
 \label{eq:implemented}
\end{equation}
{
The factor $1/8$ combines $1/2$ from averaging the squared real transfer
and $1/4$ from the retained radial exponential. The relative phase is
preserved across the finite recombination width. The time and band
averages and their conditions are described in the Supplemental
Material~\cite{ApproxSupplement}.
}

{\section{Coherent limit and scaling}}

For the additional limits $R_{\rm eff}\to-1/8$, $x\gg\sqrt L$, and
nearly constant relative phase across recombination, the amplitude
in Eq.~\eqref{eq:amplitudes} approaches $1/(2k\chi)$. Defining
$I_{\rm rec}=\int_{\rm rec}g/(a\chi)\dd\eta$ gives
\begin{equation}
 D_\ell^{BB}\simeq
 \frac{A_tLT_{\rm CMB}^2\mathcal B^2I_{\rm rec}^2}{8192\Delta}
          (k_-^{-4}-k_+^{-4}).
 \label{eq:closed}
\end{equation}
{The coefficient is derived in the Supplemental
Material~\cite{ApproxSupplement}.}
One inverse power of $k$ comes from the metric amplitude and one from
projection, giving $k^{-4}$ in power. The factor $L$ comes from the
definition of $D_\ell$; $C_\ell$ is approximately constant in this
coherent limit.

{
For a source width $\sigma_\eta$ and representative distance $\chi_*$,
phase coherence requires $L\sigma_\eta/(2k\chi_*^2)\ll1$ and a nearly
constant response phase. At higher multipoles, the finite-width
expression in Eq.~\eqref{eq:implemented} accounts for the resulting
phase cancellation. The calculation assumes linear tensor perturbations
and negligible backreaction on the background expansion.
}

\section{Validation and computational cost}
We use a fixed Planck 2018 background~\cite{Planck2018} and
$A_t=1.26\times10^{-10}$ without fitting a normalization to CLASS.
Each band has logarithmic width $\Delta=\ln10/4$, so $k_+=10^{1/4}k_-$. Starting at $k_-=10^{-1/4}\,{\rm Mpc}^{-1}$, we multiply both edges by $10^{1/4}$ to define successive bands.
We validate the approximation for five bands centered at $k_c=0.781$, $1.39$, $2.47$, $4.39$, and $7.81\,{\rm Mpc}^{-1}$, and extend the calculation to $k_c=13.9$, $24.7$, $43.9$, $78.1$, and $139\,{\rm Mpc}^{-1}$.
The recombination
integral covers $100\leq\eta\leq2000\,{\rm Mpc}$ with 4000 time points,
160 angular nodes, and 32 logarithmic-band nodes. Source responses are
reused across multipoles. {We evaluate spectra over $2\leq\ell\leq5000$.}

\begin{figure}[!htb]
 \centering
 \includegraphics[width=.64\linewidth]{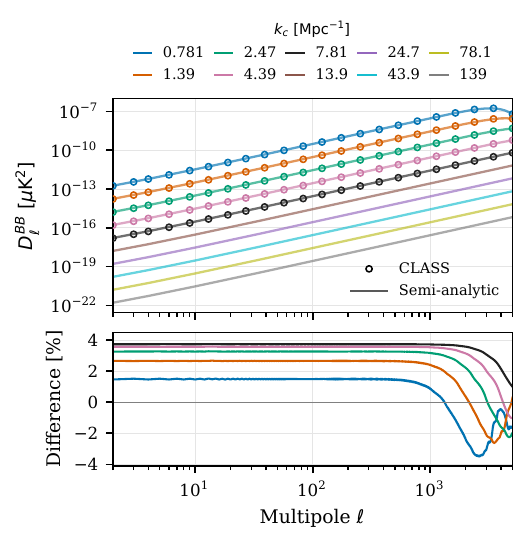}
 \caption{Narrow-band $BB$ spectra for $A_t=1.26\times10^{-10}$. Solid lines in the upper panel show semi-analytic results for ten bands with $k_c=0.781$--$139\,{\rm Mpc}^{-1}$. Open circles show refined CLASS results at selected multipoles for the five bands with $k_c=0.781$--$7.81\,{\rm Mpc}^{-1}$. Higher-wavenumber curves are semi-analytic predictions. The lower panel shows $D_{\ell,\mathrm{SA}}^{BB}/D_{\ell,\mathrm{CLASS}}^{BB}-1$ as a percentage for the five CLASS-validated bands. Colors denote $k_c$ in ${\rm Mpc}^{-1}$ in both panels.}
 \label{fig:accuracy}
\end{figure}

The upper panel of Fig.~\ref{fig:accuracy} shows the semi-analytic $BB$ spectra for all ten bands and the CLASS results for the five validation bands. The lower panel quantifies their relative differences where a CLASS reference is available. The maximum $BB$ differences are below 3.6\%
for the four bands centered at
$k_c=0.781$, $1.39$, $2.47$, and $4.39\,{\rm Mpc}^{-1}$. The corresponding
maxima below $\ell=600$ are 1.52\%, 2.67\%, 3.28\%, and 3.58\%.
The nearly constant low-$\ell$ offsets measure different asymptotic
{normalizations.} Doubling all
three quadrature resolutions changes the tested low-$\ell$ offsets
by less than 0.0005\% of the spectrum. Metric matching, local source
truncation, and discarded phases remain possible contributors; {their combined effect is assessed here.} The simpler Eq.~\eqref{eq:closed} has
maximum differences of 9.37\%, 5.14\%, 1.91\%, and 0.33\% below
$\ell=600$. Accidental cancellation of omitted terms can make the
closed expression closer in an individual band.

{For the band centered at $k_c=7.81\,{\rm Mpc}^{-1}$, the completed refined CLASS calculation gives a maximum $BB$ difference of 3.74\% over $2\leq\ell\leq5000$. Thus, the semi-analytic spectra agree with refined CLASS calculations to within 3.8\% across the five bands with $k_c=0.781$--$7.81\,{\rm Mpc}^{-1}$.} The five higher-wavenumber bands with $k_c=13.9$--$139\,{\rm Mpc}^{-1}$ have no direct CLASS reference.
{Their numerical behavior is assessed through sampled quadrature checks and convergence toward the closed limit.} The safety
condition $\sqrt L\leq0.8k_-(\eta_0-2000\,{\rm Mpc})$ keeps all selected
bands away from the turning point. The finite time interval omits a
complete late-reionization calculation. Additional convergence tests and numerical inputs are described
in Ref.~\cite{ApproxSupplement}.

\begin{table}[t]
 \caption{{CPU time in seconds for calculating the $B$-mode spectrum of each band over $2\leq\ell\leq5000$. Band centers $k_c$ are in ${\rm Mpc}^{-1}$. CLASS times are estimated from representative subbands; semi-analytic times are the mean and standard deviation of ten runs. The estimated ratio is the CLASS time divided by the semi-analytic time. Background precomputation for the semi-analytic method is excluded.}}
 \label{tab:timing}
 \begin{ruledtabular}
 \begin{tabular}{crrr}
 $k_c$ & CLASS total, est. & Semi-analytic & Est. ratio\\
 \hline
 0.781 & $4.94\times10^{3}$ & $3.22\pm0.37$ & $1.53\times10^{3}$\\
 1.39 & $1.54\times10^{4}$ & $2.63\pm0.58$ & $5.86\times10^{3}$\\
 2.47 & $5.02\times10^{4}$ & $2.80\pm0.42$ & $1.79\times10^{4}$\\
 4.39 & $1.66\times10^{5}$ & $2.87\pm0.40$ & $5.79\times10^{4}$\\
 \end{tabular}
 \end{ruledtabular}
\end{table}

{Table~\ref{tab:timing} compares single-thread CPU times for generating the $B$-mode spectra. The semi-analytic method takes 2.6--3.2 seconds per band, approximately $1/1500$--$1/58000$ of the CLASS time estimated from representative subbands. {These ratios describe spectrum generation with precomputed backgrounds for the semi-analytic method.} Measurement conditions and the estimation procedure are given in the Supplemental Material~\cite{ApproxSupplement}.}

\section{Discussion: BK15 and the extension to higher \texorpdfstring{$k$}{k}}
{
The BK15 application tests how the accuracy of the semi-analytic spectra propagates to constraints on primordial tensor power. Keeping the observational inputs and statistical assumptions fixed, we replace the CLASS spectra with the semi-analytic spectra. For the five bands with $k_c=0.781$--$7.81\,{\rm Mpc}^{-1}$, this changes the 95\% upper amplitude $A_{95}$ by 1.47--3.60\%. This agreement shows that the method reproduces the inferred amplitude bounds to a few percent in the CLASS-validated range, enabling efficient use of the spectra in observational analyses.

We use the 78 $BB$ auto- and cross-spectra in nine BK15 multipole bins, together with the released 702-dimensional covariance and bandpower windows~\cite{BK15,BK15data}. The released lensed-CMB and foreground (FG) fiducial model is fixed and subtracted from the data. Each primordial band is fitted separately, with the amplitudes of all other bands set to zero. We adopt a uniform prior on $A_t\geq0$ and define $A_{95}$ as the 95\% posterior quantile. Holding cosmological and FG parameters fixed provides a conditional analysis that isolates the effect of the spectral approximation on the bounds. FG marginalization is outside the scope of this application. The conditional Gaussian likelihood and the definition of $A_{95}$ are
given in Eqs.~(S20)--(S21) of the Supplemental Material~\cite{ApproxSupplement}.
}

{The observational calculation uses completed CLASS spectra for bands with $k_c=1.39\times10^{-4}$--$7.81\,{\rm Mpc}^{-1}$ and semi-analytic spectra for the five bands with $k_c=13.9$--$139\,{\rm Mpc}^{-1}$. In the overlap, the five bands with $k_c=0.781$--$7.81\,{\rm Mpc}^{-1}$ are evaluated with both methods using the same data and likelihood.}

\begin{figure}[!tbp]
 \centering
 \includegraphics[width=.82\linewidth]{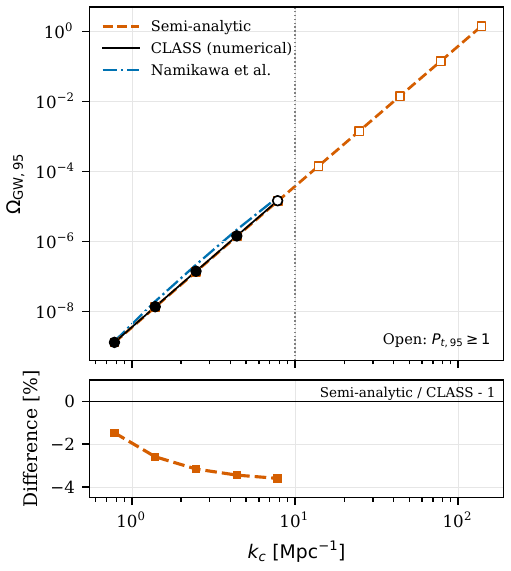}
 \caption{{Conditional BK15 95\% upper bounds. Black solid lines and circles show the numerical CLASS results; orange dashed lines and squares show the semi-analytic approximation; the blue dash-dotted line shows the $B$-mode bounds read from Fig.~4 of Namikawa et al.~\cite{Namikawa2019}. The lower panel gives the relative difference of the semi-analytic bounds from CLASS. The vertical line marks $10\,{\rm Mpc}^{-1}$. Open symbols denote formal extrapolations of the linear response with $P_{t,95}\geq1$. Our bounds use the oscillation-averaged energy conversion, while the earlier results retain their published convention.}}
 \label{fig:bk15}
\end{figure}

{
To compare with earlier GW constraints, we express the amplitude bounds as an oscillation-averaged energy density. For modes entering the horizon during radiation domination, Eq.~\eqref{eq:metric} and $\Omega_{\rm GW}=P_t|T'|^2/(12H_0^2)$ give
\begin{equation}
 \overline\Omega_{\rm GW,95}\simeq
 \frac{D_\nu^2\Omega_r}{24}\frac{A_{95}}{2\Delta}.
 \label{eq:omega}
\end{equation}
As shown in Fig.~\ref{fig:bk15}, our bounds broadly agree with those of Namikawa et al.~\cite{Namikawa2019} in magnitude and wavenumber dependence. The residual differences may reflect the different covariance treatment, foreground assumptions, and tensor-transfer conventions adopted in the two analyses.

The main benefit of the reduced computational cost is access to higher primordial wavenumbers. With the same BK15 data, we extend the calculation to five additional bands with $k_c=13.9$--$139\,{\rm Mpc}^{-1}$. In these bands, the response is predicted by the semi-analytic calculation with numerical convergence checks; direct CLASS validation covers the five bands with $k_c=0.781$--$7.81\,{\rm Mpc}^{-1}$. Ng~\cite{Ng2022} already explored this wavenumber range using CMB temperature anisotropy; here we evaluate the $B$-mode response. At fixed observed multipoles, the response decreases approximately as $k_c^{-4}$. Consequently, a larger primordial amplitude is needed to produce a comparable observed signal, and the formal amplitude bounds grow approximately as $k_c^4$.

The high-wavenumber extension also quantifies the wavenumber dependence of observational sensitivity. In the BK15 analysis, the inferred band power $P_{t,95}=A_{95}/(2\Delta)$ reaches unity by the band centered at $k_c=7.81\,{\rm Mpc}^{-1}$ and becomes $5.79\times10^5$ at $k_c=139\,{\rm Mpc}^{-1}$. The open symbols in Fig.~\ref{fig:bk15} show the formal extrapolation of the linear response in this regime. Interpreting such values as physical upper bounds requires satisfying the weak-perturbation assumption used in the derivation.

Placing these results in the broader frequency landscape clarifies the role of high-$k$ CMB constraints on $\Omega_{\rm GW}$. With $a_0=1$, the present frequency is $f=ck/(2\pi)$, so $k_c\simeq140\,{\rm Mpc}^{-1}$ corresponds to $f_c\simeq2.2\times10^{-13}\,{\rm Hz}$. Our extension maps the $B$-mode sensitivity to localized primordial power at these frequencies, including the transition to the formal linear-response bounds described above. It thus provides a low-frequency complement to pulsar timing arrays (PTAs), which probe the nanohertz background~\cite{NANOGrav2023}, and to interferometric searches: LISA targets the millihertz band~\cite{LISA2017}, the proposed CHRONOS detector targets $0.1$--$10\,{\rm Hz}$~\cite{CHRONOSCQG2026}, and LIGO--Virgo observations constrain stochastic backgrounds at tens to hundreds of hertz~\cite{LVK2021}. Together, these frequency windows test different portions of a primordial spectrum; relating their constraints requires specifying its frequency dependence and cosmological transfer history.

The method makes it practical to evaluate the $B$-mode response to small-amplitude primordial gravitational waves at wavenumbers where direct calculations are costly. Its few-percent accuracy in the CLASS-validated range, together with its computational reach to $k_c\simeq140\,{\rm Mpc}^{-1}$, provides a basis for parameter searches involving narrow primordial spectra and sensitivity forecasts for future observations.
}

\section{Conclusion}
The finite-band high-$k$ tensor response can be evaluated from local
photon envelopes and a coherent finite-width projection. Resolving the
wave--projection difference phase preserves the high-multipole shape
beyond the closed $Lk_c^{-4}$ limit. The implemented $BB$
{templates agree with refined CLASS within 3.8\% on the five tested bands through $\ell=5000$. For the four bands with $k_c=0.781$--$4.39\,{\rm Mpc}^{-1}$, the measured $B$-mode generation cost is 2.6--3.2 CPU seconds per complete band with precomputed inputs.} The BK15 demonstration is broadly consistent with earlier $BB$ limits
on their overlap. {{The higher-$k$ extension satisfies numerical convergence checks.} Direct CLASS validation covers $k_c=0.781$--$7.81\,{\rm Mpc}^{-1}$, and the high-$k$ BK15 bounds represent a formal extrapolation beyond the linear regime.} The resulting response library provides a basis
for analyses of resolved finite primordial bands, while keeping source
model error distinct from {numerical error}.

\section*{Data availability}
The BK15 data, covariance matrices, and bandpower windows used in this
work are publicly available from the BICEP/Keck
Collaboration~\cite{BK15data}. The numerical method, adopted parameters,
and validation results are described in this article and its
Supplemental Material.

\begin{acknowledgments}
We thank M.~Hazumi, H.-Y.~Huang, Y.~Mikura, T.~Okumura,
K.~W.~Ng, Y.~Chinone, T.~Namikawa, and G.~Galloni
for helpful discussions and comments.
Y.I. is supported by the National Science and Technology Council
(NSTC), CHiP in Taiwan under Grant Nos.~114-2112-M-008-006- and 115-2112-M-008 -022 -.
\end{acknowledgments}

\bibliographystyle{apsrev4-2}
\bibliography{Approximation_refs}
\end{document}


\raggedbottom
\title{\texorpdfstring{Supplemental Material: Fast finite-width CMB B modes\\
from narrow primordial spectra}{Supplemental Material: Fast finite-width CMB B modes from narrow primordial spectra}}
\author{Yuki Inoue}
\noaffiliation
\maketitle

This supplement supplies derivation steps, reproducibility information,
and the conditional BICEP/Keck 2015 (BK15) likelihood. The observable is
$BB$ throughout. {We use the notation of the main text.} The cosmic
microwave background (CMB) source retains the coupled photon distributions.

\section{Additional derivation and validity checks}
{
Our calculation builds on the analytic tensor-polarization framework of
Pritchard and Kamionkowski~\cite{Pritchard2005} and the photon-source
integral equation of Weinberg~\cite{Weinberg2006}.
Equations numbered (S1), (S2), and so on belong to this Supplemental
Material. References to equations in the main text are identified explicitly.
For external equations, we name the authors at each reference; the equation
numbers follow arXiv:astro-ph/0412581v1 for Pritchard and Kamionkowski
and arXiv:astro-ph/0607076v2 for Weinberg.

For the angular spectra, we use the line-of-sight structure of
Eqs.~(16)--(20) of Pritchard and Kamionkowski, expressed in our primordial
power and source normalization. The $B$-mode projection factor in
Pritchard and Kamionkowski, Eq.~(40), satisfies
$P_{B\ell}=4r_\ell^B$, with $r_\ell^B$ defined in Eq.~(3) of the main text. The oscillatory Bessel
approximation in Pritchard and Kamionkowski, Eq.~(41), and their
projection-envelope averages, Eqs.~(42)--(44), provide the basis for the radial asymptotic treatment. Their Eq.~(41)
uses $\ell+1/2$, whereas our WKB phase uses $\sqrt{\ell(\ell+1)}$
and matches the large-$x$ phase; these are distinct asymptotic prescriptions.
We retain the relative phase between the tensor oscillation and radial
projection inside the visibility integral. Pritchard and Kamionkowski, Eqs.~(51)--(54) and
(60), describe the source and finite-width phase damping using a
tight-coupling source approximation and an approximate visibility
average. We instead evaluate the complex local source response over the
finite visibility profile, keeping this source--projection phase.

For the source, we start from the angular Boltzmann equations in
Weinberg, Eqs.~(22)--(24). The line-of-sight solution in Weinberg,
Eq.~(28), gives the closed history integral in Weinberg, Eq.~(29),
with kernels defined in Weinberg, Eq.~(30).
These equations underlie the retarded propagator and source integral
below. Weinberg solves the history equation by iteration beginning with
Weinberg, Eq.~(31). Our high-$k$ calculation approximates the same photon response
by extracting the rapid tensor oscillation and expanding the remaining
envelope through its first time derivative, as derived below.
}

\subsection{From photon history to a local response}
For the angular photon equations (6) of the main text, the retarded
propagator is
\begin{equation}
 K_\mu(\eta,\eta')=\exp\!\left[-\int_{\eta'}^\eta q(s)\dd s
                            -\ii k\mu(\eta-\eta')\right].
\end{equation}
Here $q>0$ is the conformal Thomson rate and $\mu$ is the photon-direction
cosine. Integrating the angular equations, then applying the weights
$w_F,w_G$ defined in Eq.~(7) of the main text, gives~\cite{Weinberg2006}
\begin{align}
 p(\eta)={}&p_{\rm hom}(\eta)
  -\int_{\eta_i}^\eta e^{-Q}\mathcal K_F(z)T'(\eta')\dd\eta'\notag\\
 &+\int_{\eta_i}^\eta e^{-Q}q(\eta')p(\eta')
       [\mathcal K_F(z)+\mathcal K_G(z)]\dd\eta'.
\end{align}
{The first integral describes the metric driving, while the
second contains the scattering feedback through $p$. Their angular
kernels are}
\begin{align}
 \mathcal K_F(z)&=\frac32\frac{j_2(z)}{z^2},\notag\\
 \mathcal K_F(z)+\mathcal K_G(z)
 &=\frac32\left[j_0(z)-\frac{2j_1(z)}z+\frac{2j_2(z)}{z^2}\right].
\end{align}
In these expressions $Q=\int_{\eta'}^\eta q\dd s$,
$z=k(\eta-\eta')$, $T$ is the metric transfer, and $p_{\rm hom}$ is the
source projection of the initial photon distributions at $\eta_i$.
The continuous kernel values at $z=0$ are $1/10$ and $7/10$.
In the convention of Ref.~\cite{Weinberg2006}, our variables satisfy $F=-\Delta_T^{(T)}/2$,
$G=-\Delta_P^{(T)}/2$, and $p=\Psi/2$, with the same primordial
metric normalization. Weinberg's $\Psi$ is a source amplitude,
whereas our $\Psi_\ell$ is a projection phase.
{With $\dd t=a\,\dd\eta$ and $q(\eta)=a\omega_c$, Weinberg's
wave number corresponds to our $k$. In terms of his kernels $K$ and $F$
in Weinberg, Eq.~(30), our kernels obey
$\mathcal K_F=3K/2$ and $\mathcal K_F+\mathcal K_G=3F/2$.
The term $p_{\rm hom}$ retains the initial photon contribution explicitly;
The early equilibrium initial condition is stated immediately after
Weinberg's Eq.~(28).}

{This integral equation retains the photon memory, which we approximate below by a local response.}

{
For the positive-frequency metric branch, we factor
$(\mathcal B/k)e^{\ii(k\eta+\varphi)}$ from the photon distributions.
The negative-frequency source follows by complex conjugation and the
symmetry $\mu\mapsto-\mu$ of the angular weights. The derivative of the
extracted factor contributes $\ii k$, so the envelope coefficient is
$d_\mu(\eta)=q(\eta)+\ii k(1+\mu)$.

We first consider a generic envelope $Y$ driven by a prescribed source
$b$. At fixed $k$ and $\mu$, its equation is
\begin{equation}
 (d_\mu+\partial_\eta)Y
 \equiv d_\mu(\eta)Y(\eta)+Y'(\eta)=b(\eta).
 \label{eq:envelopeODE}
\end{equation}
Here a prime denotes a conformal-time derivative: $d_\mu$ acts by
multiplication, whereas $\partial_\eta$ differentiates the envelope.
Since $q>0$, $d_\mu$ is nonzero, and Eq.~\eqref{eq:envelopeODE} of this Supplemental Material can be
rewritten as $Y=b/d_\mu-Y'/d_\mu$. We obtain the leading local response
by omitting the envelope derivative, and then substitute that response
into the derivative term:
\begin{equation}
 Y_0=\frac{b}{d_\mu}.
\end{equation}
The time variation of this leading response supplies the first correction,
\begin{equation}
 Y_1=-\frac{Y_0'}{d_\mu}
     =-\frac1{d_\mu}\frac{\dd}{\dd\eta}
                        \left(\frac{b}{d_\mu}\right).
\end{equation}
Thus the first-order local solution is
\begin{equation}
 Y\simeq Y_0+Y_1
   =\frac{b}{d_\mu}-\frac{b'}{d_\mu^2}
                    +\frac{b\,d_\mu'}{d_\mu^3}.
 \label{eq:inverse}
\end{equation}
The last term accounts for the time dependence of the scattering rate,
with $d_\mu'=q'$. Multiplication by $d_\mu$ and time differentiation
therefore do not commute. Substitution into the original equation gives
\begin{equation}
 (d_\mu+\partial_\eta)(Y_0+Y_1)=b+Y_1',
\end{equation}
so the remaining term is of second order in envelope derivatives.
This construction is the local derivative expansion denoted by
$(d_\mu+\partial_\eta)^{-1}b$; the initial-condition contribution is
carried separately by the homogeneous solution.

For the coupled photon envelopes, the forcing also contains the
complex polarization-source envelope $p_0+p_1+\cdots$. We normalize the leading complex source envelope $p_0$ and its first
correction $p_1$ by the metric envelope $U$, as in Eq.~(9) of the main text:
\begin{equation}
 R_{\rm eff}=(p_0+p_1)/U.\label{eq:s-response}
\end{equation}
Restoring the extracted oscillation and taking the real part gives the
physical polarization source, reproducing Eq.~(10) of the main text,
\begin{equation}
 p(k,\eta)\simeq\frac{\mathcal B}{k}
 \operatorname{Re}\!\left[U R_{\rm eff}e^{\ii(k\eta+\varphi)}\right].\label{eq:s-source}
\end{equation} With the main text's notation
$d=\ii kU+U'$, the leading photon equations form the coupled pair
\begin{equation}
 d_\mu F_0=d-qp_0,\qquad d_\mu G_0=qp_0.
\end{equation}
At first order, derivatives of these leading envelopes provide additional
driving terms, while $p_1$ describes the corresponding source correction:
\begin{equation}
 d_\mu F_1=-qp_1-F_0',\qquad d_\mu G_1=qp_1-G_0'.
\end{equation}
The terms proportional to $p_1$ enforce the same angular source relation
at first order. Applying the weights $w_F,w_G$ gives the main text's
$p_1$ and hence $R_{\rm eff}=(p_0+p_1)/U$.
}
Thus the reduction truncates an envelope-derivative
{series and discards the attenuated homogeneous transient while retaining the finite visibility width.} Away from envelope zeros,
sufficient local diagnostics are
\begin{equation}
 \frac{|F_0'|}{|d_\mu F_0|},\quad
 \frac{|G_0'|}{|d_\mu G_0|},\quad
 \frac{|d_\mu'|}{|d_\mu|^2}\ll1.
\end{equation}
At a zero, absolute successive source contributions must be compared.
{Since $d_{-1}=q$, the endpoint expansion is controlled by the scattering rate.}

The extra limit needed only for the closed coefficient can be checked
without calibration. {We set} $\delta=q/k$, $j_{F,G}=kJ_{F,G}$, discard $U'$
and the derivative correction, and write
$R=-\ii j_F/[1-\delta(j_F+j_G)]$. Here $R=p_0/U$ is
the leading, slowly varying-envelope limit of $R_{\rm eff}$;
$J_F$, $J_G$, and $M$ are defined in Eq.~(8) of the main text.
{As $\delta\to0$, the angular weight in $j_F$ cancels the
endpoint factor in the denominator, giving the finite limit}
\begin{equation}
 j_F\longrightarrow\frac{3}{32\ii}
       \int_{-1}^1(1-\mu)^2(1+\mu)\dd\mu=\frac1{8\ii}.
\end{equation}
{For $j_G$, the endpoint weight remains nonzero. We therefore
bound its contribution to the response denominator before taking the limit:}
\begin{equation}
 |\delta j_G|\leq\frac{w_{G,\max}}2\,
       \delta\operatorname{arsinh}(2/\delta)\longrightarrow0,
\end{equation}
where $w_{G,\max}=\max_{[-1,1]}w_G$.
{Since $j_F$ is finite, $\delta j_F$ also vanishes, so the
response denominator tends to unity and $R\to-1/8$.}
The second estimate treats the nonvanishing endpoint weight before
taking the limit. This fast-wave limit $q/k\ll1$ is distinct from tight
coupling; scattering still enters through the visibility $g=q e^{-\tau}$.

\subsection{Discarded radial terms and the two phase operations}
With $b_\ell=(x\sqrt{\mu_\ell})^{-1}$, differentiation of the main
text's Wentzel--Kramers--Brillouin (WKB) form, Eq.~(11), gives
\begin{equation}
 r_\ell^B\simeq\frac{b_\ell\mu_\ell}{2}\cos\Phi_\ell
 +\frac12\left(\frac{\dd b_\ell}{\dd x}
                     +\frac{2b_\ell}{x}\right)\sin\Phi_\ell.
\end{equation}
Only the first term is retained. {The relative size of the terms is assessed using their envelopes.} The diagnostic
$L/(x^3\mu_\ell^3)$ fails near the turning point.

{
The product of the source oscillation and radial projection contains
sum- and difference-phase terms. When the rapidly varying sum-phase
contribution is suppressed by integration over the smooth recombination
profile, we retain the relative phase $\Psi_\ell=\Phi_\ell-x$ inside the
time integral. Separately, averaging the squared real transfer over a
primordial band gives a factor of $1/2$ when the overall phase varies
sufficiently across the band and the complex response varies slowly
on that oscillation scale. This band average preserves the relative
phase within the recombination integral. The CLASS comparison in the
main text assesses the combined accuracy of these phase treatments
and the source and radial approximations.
}

{
\subsection{Derivation of the coherent limit}
The local-response denominator satisfies $\operatorname{Re}M\geq3/10$
for $q>0$ and therefore remains nonzero.

The coherent limit of the main text assumes $R_{\rm eff}\simeq-1/8$,
$k\chi\gg\sqrt{L}$, and a nearly constant relative phase across
recombination. Substituting $U=1/a$ and
$\mathcal A_B\simeq1/(2k\chi)$ into the main text's definition of
$\mathcal J_B$ in Eq.~(14) separates the constant phase and wavenumber dependence
from the remaining recombination integral. We denote the latter by
\begin{equation*}
 I_{\rm rec}=\int_{\rm rec}\frac{g}{a\chi}\dd\eta.
\end{equation*}
This integral retains the finite recombination width. The coherent
projection amplitude then takes the form
\begin{equation}
 \mathcal J_B\simeq-\frac{e^{\ii\Psi_{\ell,*}}}{16k}I_{\rm rec}.\label{eq:s-coherentJ}
\end{equation}
Here $\Psi_{\ell,*}$ is the approximately constant value of the
relative phase within the recombination layer.
Using the band average
$\langle k^{-4}\rangle_k=(k_-^{-4}-k_+^{-4})/(4\Delta)$,
with $\Delta=\ln(k_+/k_-)$, in the main text's expression for
$D_\ell^{BB}$, Eq.~(15), yields
\begin{equation}
 D_\ell^{BB}\simeq
 \frac{A_tLT_{\rm CMB}^2\mathcal B^2 I_{\rm rec}^2}{8192\Delta}
 (k_-^{-4}-k_+^{-4}).
\end{equation}
This reproduces the normalization of Eq.~(16) of the main text.
}

\section{Numerical inputs and supplementary checks}
Numerical inputs are collected in Table~\ref{tab:parameters}.
\setlength{\intextsep}{6pt}
\begin{table}[htb]
\caption{Fixed $\Lambda$CDM
background~\cite{Planck2018} and source normalizations. The first four
rows are inputs;
$\eta_0$, $H_0$, $\Omega_r$, $\mathcal B^2$, and $I_{\rm rec}$ are derived.}
\label{tab:parameters}
\renewcommand{\arraystretch}{0.94}
\begin{ruledtabular}
\begin{tabular}{ll}
Parameter & Value\\
\hline
$h$ & $0.6736$\\
$\omega_b=\Omega_b h^2$ & $0.02237$\\
$\omega_c=\Omega_c h^2$ & $0.120$\\
$\tau_{\rm reio}$ & $0.0544$\\
$\eta_0$ & $14174.5563\,{\rm Mpc}$\\
{$H_0$} & {$67.36\,\mathrm{km\,s^{-1}\,Mpc^{-1}}$}\\
$\Omega_r$ & $9.218085625\times10^{-5}$\\
$D_\nu$ & $0.8016386884$~\cite{Weinberg2004}\\
$T_{\rm CMB}$ & $2.7255\,{\rm K}$~\cite{Fixsen2009}\\
$\mathcal B^2=D_\nu^2H_0^2\Omega_r$ & $2.990617199\times10^{-12}\,{\rm Mpc}^{-2}$\\
$I_{\rm rec}$ & $0.0709456149\,{\rm Mpc}^{-1}$\\
\end{tabular}
\end{ruledtabular}
\par\smallskip
\footnotesize\raggedright
The displayed $H_0$ uses conventional velocity units. In
$\mathcal B^2$ and the transfer equations, $H_0$ denotes
$H_0/c=2.2469\times10^{-4}\,{\rm Mpc}^{-1}$, as in the main text.
$\omega_{b,c}$ are physical baryon/cold-dark-matter densities, and
$\tau_{\rm reio}$ is reionization optical depth.
$I_{\rm rec}=\int_{\rm rec}g/(a\chi)\,\dd\eta$ uses the stated interval.
{The displayed digits specify the numerical settings;}
the BK15 fiducial is a separate experimental input.
\end{table}

The numerical resolution and radial safety condition are stated in the
main text. At the most restrictive point, $k_-=10^{-1/4}\,{\rm Mpc}^{-1}$ at the lower edge of the band centered at $k_c=0.781\,{\rm Mpc}^{-1}$ and
$\ell=5000$, $\sqrt L/[k(\eta_0-2000\,{\rm Mpc})]=0.730$ and
$L/(x^3\mu_\ell^3)\simeq2.4\times10^{-4}$. These numbers check radial
validity only. {For the band centered at $k_c=7.81\,{\rm Mpc}^{-1}$, the refined reference covers the complete band. The maximum change in $BB$ between the coarse and refined calculations is 0.24\% under the convergence criterion used here.} 
{We denote by $N_\eta$ the number of uniformly spaced conformal-time points used for trapezoidal integration over $100\leq\eta/\mathrm{Mpc}\leq2000$. The quantities $N_\mu$ and $N_k$ are the numbers of Gauss--Legendre quadrature nodes for the photon-direction cosine $\mu=\hat{\boldsymbol k}\cdot\hat{\boldsymbol n}\in[-1,1]$ and for $\ln k\in[\ln k_-,\ln k_+]$ within each primordial band, respectively. Thus, $N_k$ counts integration nodes within a band, and $\mu$ denotes the photon direction rather than the radial WKB quantity $\mu_\ell$.}

Doubling $(N_\eta,N_\mu,N_k)$ from $(4000,160,32)$ to
$(8000,320,64)$ at $\ell=2,80,300,600,2000,5000$ changes $BB$ by at
most 0.0041\% and 0.0039\% at $k_c=0.781$ and $4.39\,{\rm Mpc}^{-1}$, respectively, and below 0.0005\% in the six bands with $k_c=7.81$--$139\,{\rm Mpc}^{-1}$. {These tests assess numerical convergence in the five extension bands centered at $k_c=13.9$, $24.7$, $43.9$, $78.1$, and $139\,{\rm Mpc}^{-1}$; direct CLASS validation covers the five bands with $k_c=0.781$--$7.81\,{\rm Mpc}^{-1}$.}

{\section{Computational cost comparison}}
{
We compare the CPU time required by CLASS and the semi-analytic method to calculate the $B$-mode spectrum of each band over $2\leq\ell\leq5000$. Measurements were performed on an Apple M5 MacBook Pro with 16 GB of memory, with both methods run sequentially using a single thread.

For CLASS, we divide the subband list for each band into ten intervals and time the central subband of each interval. Denoting these CPU times by $t_j$ and the total number of subbands by $N_{\rm sub}$, we estimate the complete-band cost as
\begin{equation}
 \widehat T_{\rm CLASS}=\frac{N_{\rm sub}}{10}\sum_{j=1}^{10}t_j.
\end{equation}
The counts $N_{\rm sub}=147,411,1207,3654$ correspond to the band centers $k_c=0.781$, $1.39$, $2.47$, and $4.39\,{\rm Mpc}^{-1}$, respectively. For the semi-analytic method, we calculate each complete band ten times and report the mean CPU time and sample standard deviation. We also verify that the spectra used for timing are consistent with those used for accuracy validation.

The measurements include input reading through spectrum output, but exclude background precomputation for the semi-analytic method and the final summation of CLASS subbands. Since CLASS calculates the background separately for each subband, the time ratios in the main text apply to this implementation and these measurement conditions.
}

\section{Conditional BK15 amplitude likelihood}
\label{sec:bk15-likelihood}
For each primordial band we form the 702-component residual
$\boldsymbol y$ by subtracting the released fixed lensed-CMB and
foreground fiducial from the 78 measured $BB$ spectra in nine
multipole bins~\cite{BK15,BK15data}. The covariance $\mathbf C$ is
retained without subtraction. Let $\boldsymbol t$ be the corresponding
bandpower vector obtained by applying the released windows to the tensor
spectrum at unit $A_t$, using the normalization of Eq.~(1) of the main text.
The conditional Gaussian likelihood is
\begin{equation}
 -2\ln\mathcal L(A_t)
 =(\boldsymbol y-A_t\boldsymbol t)^T\mathbf C^{-1}
 (\boldsymbol y-A_t\boldsymbol t)+\mathrm{const.}
 \label{eq:bk15-like}
\end{equation}
The scalar cosmology and foreground model are fixed, and all other
primordial bands have zero amplitude. This is the conditional Gaussian
application used here, rather than the collaboration's full foreground-
marginalized likelihood. With a uniform prior on $A_t\geq0$, the upper
amplitude $A_{95}$ is defined by
\begin{equation}
 \frac{\int_0^{A_{95}}\mathcal L(A_t)\dd A_t}
 {\int_0^\infty\mathcal L(A_t)\dd A_t}=0.95.
 \label{eq:bk15-quantile}
\end{equation}
The same residuals, covariance, windows, and prior are used for CLASS
and the semi-analytic spectra. Only the tensor response
$\boldsymbol t$ changes in this comparison.

\bibliographystyle{apsrev4-2}
\interlinepenalty=10000
\bibliography{Approximation_refs}